# Topological descriptor for interpretable thermal transport prediction in amorphous graphene

Kosuke Yamazaki[1], Takuma Shiga[2], Kumpei Shiraishi[1,3], Emi Minamitani*[1,3]

[1] *Graduate School of Engineering Science, The University of Osaka, 1-3, Machikaneyama, Toyonaka, Osaka, Japan*

[2] *Department of Advanced Science and Technology, Toyota Technological Institute, 2-12-1 Hisakata, Tempaku, Nagoya, Aichi, 468-8511, Japan*

[3] *SANKEN, The University of Osaka, Ibaraki, Osaka, 565-0871, Japan*

**Abstract**

Understanding and predicting thermal transport in disordered materials remains a significant challenge due to the absence of periodicity and the complex nature of medium-range structural motifs. In this work, we investigate amorphous graphene and demonstrate that persistent homology, a topological data analysis technique, can serve as a physically interpretable structural descriptor for predicting thermal conductivity.

We first show that ridge regression using persistent homology descriptors achieves high prediction accuracy. To gain physical insight into the prediction process, we perform inverse analysis by mapping the regression coefficients back onto the persistence diagrams. This reveals that distorted hexagonal and triangular motifs are strongly correlated with reduced thermal conductivity. A further comparison with the spatial distribution of localized vibrational modes supports the physical interpretation that these motifs suppress thermal transport.

Our findings highlight that persistent homology not only enables accurate physical property prediction but also uncovers meaningful structure–property relationships in two-dimensional amorphous materials. This approach offers a promising framework for interpretable machine-learning models in materials science.

**Introduction**

Amorphous materials have been widely applied in practical devices such as thin-film solar cells based on amorphous silicon, oxide-semiconductor-based flat-panel displays. While most applications to date have focused on three-dimensional bulk materials, recent attention has turned to two-dimensional (2D) amorphous materials[1–3]. In particular, the successful fabrication of large-area 2D amorphous graphene (a-Gr) by Toh et al. via chemical vapor deposition has drawn considerable interest[4]. The obtained a-Gr exhibits high fracture strength and stability, as well as unique structural features not observed in crystalline counterparts[1,5–10], making it a promising candidate for next-generation electronic and energy materials.

In these two-dimensional amorphous materials, elucidating the relationship between structure and physical properties is crucial for designing higher-performance materials. Unlike crystalline solids, amorphous materials lack long-range order or periodicity in their atomic arrangements. Instead, they exhibit structural features known as medium-range order (MRO) on a scale of approximately 5 to 20 Å[11]. However, conventional structural analysis methods such as the radial distribution function (RDF) have difficulty quantitatively capturing such medium-range order as they are limited to two-point correlations.

To address this challenge, topological data analysis, particularly persistent homology (PH), has attracted growing attention [12,13]. PH is a mathematical technique that captures topological features such as "holes" and "voids" in data. When applied to atomic coordinates, it enables the extraction of structural features in a quantitative manner. For example, it has been reported that

PH can distinguish between the liquid, glass, and crystalline phases of $SiO_2$ [14].

Figure 1 illustrates the basic procedure for computing persistent homology from atomic coordinates. Virtual spheres are placed at each atomic site and gradually expanded. When two spheres come into contact, an edge is introduced between the corresponding atoms. As shown in Fig. 1b, a ring (i.e., one-dimensional hole) is formed at certain radii. With further expansion, this ring becomes entirely covered by spheres, as in Fig. 1d, and is considered to have vanished. As the radius continues to grow, larger rings are formed and then disappear (Fig. 1c-e). Throughout this process, the birth and death radii of such topological features (holes and voids) are recorded, and the result is represented as a two-dimensional diagram called the persistence diagram (PD) (Fig. 1f). The PD compactly encodes topological features across multiple scales and enables efficient visualization and quantification of structural differences in atomic configurations.

Leveraging these advantages, recent studies have explored the use of persistent homology as a structural descriptor[15–24]. For example, the atomic structures of three-dimensional amorphous semiconductors have been analyzed using persistent homology and their thermal conductivity successfully predicted with high accuracy based on topological features[25,26]. However, it remains unclear whether this structure–property prediction approach using persistent homology is applicable to two-dimensional amorphous materials such as a-Gr. Notably, two-dimensional systems allow for a direct visualization of vibrational modes, which facilitates the interpretation of how topological features relate to local physical properties and provides deeper insight into which structural characteristics govern thermal transport.

In this study, we construct a machine learning model to predict thermal conductivity of a-Gr using structural features extracted via PH. Furthermore, we propose a framework for interpreting the physical basis of the model's predictions through inverse analysis of the PH features, which provides physically meaningful insight into the structure–property relationship.

Our PH-based descriptor outperforms a conventional descriptor based on ring statistics in terms of prediction accuracy. The inverse analysis reveals that the birth–death pairs in PD most strongly associated with thermal conductivity correspond to hexagonal rings and internal triangular motifs formed by carbon atoms. The spatial distribution of these hexagonal and triangular structures, which contribute negatively to thermal conductivity, overlaps with regions of localized low-frequency vibrational modes. This demonstrates that the predictions made by PH-based models are grounded in physically relevant correlations.

This work not only demonstrates accurate prediction of thermal conductivity for two-dimensional amorphous materials using PH-based descriptors, but also shows that these predictions can be physically interpreted. Our approach provides new insights into structure–property relationships in amorphous materials, paving the way for future material design and for optimization of synthesis conditions.

## 2. Methods

### 2.1 Molecular Dynamics Simulations

To generate amorphous graphene (a-Gr) structures, we employed classical molecular dynamics (MD) simulations using the LAMMPS package[27]. The initial configuration was constructed by placing 5,929 carbon atoms in a square lattice with a lattice constant of 1.63 Å within a rectangular cell of dimensions 125 Å × 125 Å × 42 Å. Interatomic interactions were modeled using the AIREBO potential[28].

The amorphous structures were prepared via a four-step MD procedure: (i) melting, (ii) high-temperature equilibration, (iii) quenching, and (iv) room-temperature equilibration. First, the system was heated from 300 K to 10,300 K at a rate of 50 K/ps under an NVT ensemble [29,30], with all atoms constrained to move within the two-dimensional plane. After equilibration at 10,300 K for 1 ns, the system was quenched to 300 K at 19 different cooling rates ranging from 10 K/ps to 1,000 K/ps. Following the quench, the planar constraint was removed to allow motion in the out-of-plane direction, and the system was equilibrated again at 300 K for 1 ns. For each cooling rate, we performed 28 independent simulations with different initial velocities.

In total, 509 a-Gr structures with varying degrees of disorder were generated through this procedure. A representative example of the a-Gr structures obtained in this study is shown in Figure 2.

**2.2 Persistent Homology**

In this study, we used the HomCloud[31,32] library to compute persistence diagrams (PDs). To characterize the two-dimensional amorphous structures, we employed first homology groups, which capture ring-like features formed by atomic configurations. The birth radius corresponds to the maximum interatomic distance required to form a ring, while the death radius reflects the inradius of the formed ring structure. To assign the ring features associated with each birth–death pair in the PD, we employed volume-optimal cycles[33] implemented in the HomCloud library.

**2.3 Persistence Image**

While PDs are powerful tools for quantifying topological information, the number of birth–death pairs varies across structures, making them difficult to use directly as input for machine learning. To address this issue, we employed the persistence image (PI) method [34], which converts a PD into a fixed-dimensional numerical vector.

In this method, a Gaussian kernel is applied to each birth–death point $(b_k, d_k)$ to construct a continuous function $K((x, y), (b_k, d_k))$:

$$K((x, y), (b_k, d_k)) = \exp\left(-\frac{(b_k - x)^2 + (d_k - y)^2}{2\sigma^2}\right). \tag{1}$$

Here, $(x, y)$ are grid coordinates, $(b_k, d_k)$ are the coordinates of a birth–death point, and $\sigma$ is a parameter controlling the spread of the kernel. In this study, we set $\sigma=0.063$, and divided the region $(\text{birth}, \text{death}) \in ([0,4], [0,4])$ into a grid of 299 points. These parameters were determined through a parameter search to optimize prediction accuracy. To reflect the importance of ring structures based on their distance from the diagonal, we introduce a weight function as follows:

$$\omega(b_k, d_k) = 1. \tag{2}$$

Finally, the persistence image value $\rho(x_i, y_j)$ at each grid point $(x_i, y_j)$ is computed as:

$$\rho(x_i, y_j) = \sum_{k=1}^{l} \omega(b_k, d_k) \exp\left(-\frac{(b_k - x_i)^2 + (d_k - y_i)^2}{2\sigma^2}\right) \tag{3}$$

In this study, the vector composed of persistence image values is used as an input for a ridge regression model.

**2.4 Thermal Conductivity Simulations**

In this study, we evaluated the thermal conductivity of a-Gr structures using the Allen–Feldman (AF) theory[35,36]. The AF theory defines diffusivity based on the eigenmode decomposition of vibrational modes and the coupling strength between modes. Compared to the Green–Kubo method or non-equilibrium MD, it is computationally less demanding, making it suitable for statistically comparing thermal conductivities across many different structures, as in this study. Although the AF theory does not consider anharmonic effects, it has been shown to yield reliable simulation results for covalently bonded materials. For example, in the analysis of amorphous silicon by Feldman et al., the theoretical results based on the AF model converged to the experimental result at high temperatures[37]. Since amorphous graphene consists of strong covalent bonds formed by sp² hybridization, applying the AF theory is physically justified in this case. Therefore, we adopted the AF theory in this study to evaluate the thermal conductivity of a-Gr.

In the AF theory, thermal conductivity is calculated using the following expression:

$$\kappa = \frac{1}{V} \sum_i C_i(T) D_i, \qquad (4)$$

where $C_i(T)$ is the specific heat of vibrational mode $i$ at temperature $T$, and $D_i$ is the diffusivity of this mode, given by:

$$D_i = \frac{\pi V^2}{3\hbar^2 \omega_i^2} \sum_{i \neq j} |S_{ij}|^2 \delta(\omega_i - \omega_j), \qquad (5)$$

where, $S_{ij}$ is mode decomposed heat flux operator and $\omega_i$ is the frequency of mode $i$. $V$ is the volume of the system.

The diffusivity $D_i$ represents the extent to which mode $i$ contributes to heat transport through its coupling with other modes and the energy differences involved. In this study, we evaluated $D_i$ by using a custom Python package. The code is available on GitHub[38]. In numerical calculations, the delta function in Equation (5) is approximated using a Gaussian function, with the width set to five times the average spacing between eigenvalues [16,25,39].

**2.5 Ridge regression**

Ridge regression using persistence image vectors was conducted using the Scikit-learn package[40]. For ridge regression, an intercept term was included, and the input vectors constructed from the persistent image were scaled by the maximum in the dataset. The intercept term was -0.5142. The parameter for L2 norm regularization was set to 0.7 to optimize the coefficient of determination, $R^2$ and root mean-squared error (RMSE). Because the elements of the persistence image vector are non-negative, the scaling by maximum did not change the sign of the coefficient. This enabled a straightforward interpretation of the coefficient of regression.

**3. Results and discussion**

**3.1 Prediction of Thermal Conductivity based on Persistent Homology**

We investigated the relationship between cooling rate and thermal conductivity using simulation data from 509 structures. As shown in Figure 3, The thermal conductivity of a-Gr strongly depends on the cooling rate during sample preparation. When the cooling rate was slow (10 K/ps), the thermal conductivity was relatively high. As the cooling rate increased, the thermal conductivity sharply decreased and converged to an approximately constant value (~10 W/mK) above around 700 K/ps. This trend suggests that increased structural disorder due to faster cooling contributes to the suppression of thermal transport. The obtained thermal conductivity in highly disordered cases is consistent with previous reports[5–7].

Using this thermal conductivity data, we constructed a ridge regression model based on structural descriptors derived from persistent homology and persistence image. The 509 structures in the dataset were split into 407 for training and 102 for testing. We performed five random splits and conducted cross-validation. As a result, the model achieved a high prediction accuracy with an $R^2$ value of 0.983 and an RMSE of 0.642 W/mK on the test data (Figure 4). Both training and test datasets showed a strong correlation between predicted and measured values, demonstrating that persistent homology-based features are highly effective for predicting the thermal conductivity of a-Gr.

**3.2 Inverse Analysis and Feature Extraction of Structural Contributions**

The coefficients of the trained ridge regression model correspond to the contribution of each grid point in the persistence image to the thermal conductivity. Based on these coefficients, we created a colormap over the PD (Figure 5). Regions where the absolute value of the coefficient is large correspond to areas in the diagram where birth–death pairs strongly affect the thermal conductivity. We extracted regions where the coefficient is greater than 0.3 or less than −0.1, and identified the atomic structures corresponding to the birth–death pairs located in those regions. In what follows, we refer to the region with coefficients $\geq 0.3$, which contributes to higher thermal conductivity, as the "red region," and the region with coefficients $\leq -0.1$, which contributes to lower thermal conductivity, as the "blue region."

The number of birth–death pairs located in the red and blue regions varies significantly with the cooling rate. For instance, at a cooling rate of 10 K/ps, there were 1,045 pairs in the red region and 491 in the blue region, while at 1,000 K/ps, these numbers were 191 and 491, respectively. This result indicates that rapid cooling (1,000 K/ps) reduces the presence of structures that enhance thermal conductivity and increases the proportion of those that suppress it. Since a-Gr generated at 1,000 K/ps indeed exhibits low thermal conductivity, this trend is physically reasonable.

We also compared histograms of bond lengths within the ring structures formed by atoms corresponding to the identified birth–death pairs (Figure 6). The structures associated with birth–death pairs in the red region showed sharp peaks in the histogram, while those associated with the pairs in blue region exhibited broader distributions. This is attributed to the general tendency that more ordered structures contribute to higher thermal conductivity, while more disordered structures contribute to lower conductivity.

The ring structures identified through this inverse analysis were primarily hexagons and triangles. Structures that enhanced thermal conductivity were near-perfect hexagons or triangles, whereas those that reduced thermal conductivity were found to be distorted hexagons or triangles (Figure 7). The number of each type of structure is summarized in Table 1, with the high prevalence of triangular structures being particularly noteworthy.

In addition to the PH analysis, we carried out ring statistics analysis on the same structures. The threshold for C-C bond was set to 1.6 Å and King's shortest path criterion was employed to define the ring[41,42]. The results of ring size distribution in the samples generated at the cooling rate of 10 K/ps and 1,000 K/ps are summarized in Figure 8. In the slow cooling rate sample, the dominant ring structure is six-vertices ring structure. As the cooling rate become rapid, the ratio of five- and seven-vertices rings becomes increased. However, there are no three-vertices ring structure in both cases.

These results indicate that the triangles observed in the PH analysis are not physical three-membered rings constructed by nearest-neighbor C–C bonds. Instead, they appear as *children*, a term widely used in PH-based structural analysis. In a filtration process, when a larger ring (e.g., a hexagon) is formed, transient smaller cycles can momentarily appear inside it. These short-lived sub-cycles are called *children*, and their birth and death occur entirely during the formation and closure of a larger "parent" ring [14,25,43]. In our analysis, 59% of all triangle-containing birth–death pairs in both red and blue regions were identified as children of hexagonal rings, confirming that most of these triangles are topological artifacts of the filtration rather than genuine three-membered rings.

The schematic visualization of the relationship between birth and death radii of hexagonal parent and triangle child are shown in Figure 9. The birth radius determined by the half of the maximum edge length in the ring structure and the death radius corresponds to the inradius of ring structure. Because of this definition, the hexagon includes larger disorder of the edge length has larger birth radius and shorter death radius. The children triangles also reflect the order and disorder of the parental hexagonal rings, and their birth and death radii become shorter in disordered case compared to the ordered case (Figure 9). Similar tendency is clearly confirmed in the birth and death radius distributions summarized in Figure 10.

These results demonstrate that persistent homology is capable of capturing structural features beyond simple ring connectivity. As shown in the Supplementary Materials, a predictive model based solely on ring statistics shows deviation from the ground truth in high thermal conductivity regimes even if the model returns relatively high accuracy. We believe that taking children into account corresponds to extracting hierarchical structural features embedded in atomic configurations, which is a key factor contributing to the high predictive accuracy of our method.

**3.3 Physical Interpretation of the Inverse Analyses Results**

We further examined the physical significance of the structures extracted via inverse analysis. Previous studies have reported that highly localized vibrational structures hinder propagon propagation, leading to reduced thermal conductivity [14]. In our study, analysis of the frequency-resolved contributions to thermal conductivity revealed that vibrational modes below 15 THz account for approximately 80% of the total thermal conductivity (Figure 11a).

We also estimated the participation ratio (PR) of the vibrational modes in the structures generated at the cooling rates of 10 K/ps and 1,000 K/ps. A larger (smaller) PR value corresponds to a more delocalized (localized) vibrational mode. As shown in Figure 11b, the vibrational modes in the structure generated at 1,000 K/ps exhibit stronger localization.

Considering these results, we focused on the low-frequency localized modes whose frequency and PR values are below 15 THz and 0.1, respectively. We evaluated the atomic displacement amplitudes from the eigenvectors and normalized them by the maximum displacement within each mode. We then selected the atoms whose displacement amplitudes were within the top 30% for each mode. After collecting all atoms satisfying this criterion, we evaluated the spatial overlap between these atoms and the blue regions identified by the PH-based inverse analysis. Figures 11c and d visualize the spatial correlation between these two regions for 10 K/ps and 1,000 K/ps, respectively.

To quantitatively evaluate the spatial overlap between the distribution of the localized modes and the blue regions extracted from the PH-based inverse analysis, we employed statistical measures commonly used in classification problems: True Positive (TP), True Negative (TN), False Positive (FP), and False Negative (FN). Here, TP corresponds to the number of atoms located in the blue regions that also contribute to the localized modes. TN corresponds to the number of atoms that are not located in the blue regions and do not contribute to the localized modes. FP corresponds to the number of atoms that are located in the blue regions but do not contribute to the localized modes. FN corresponds to the number of atoms that are not located in the blue regions but contribute to the localized modes.

We evaluate the Accuracy, True Positive Rate (TPR), and False Positive Rate (FPR), which are defined as follows:

$$Accuracy = \frac{TP + TN}{TP + TN + FP + FN},  \quad (6)$$

$$TPR = \frac{TP}{TP + FN}, \quad (7)$$

$$FPR = \frac{FP}{TN + FP}. \quad (8)$$

The calculation results of the above three factors for the structures generated at 10 K/ps and 1,000 K/ps cooling rates are summarized in Table 2.

In both cases, the accuracy values exceed 0.8, and the TPR is much higher than the FPR. Such high accuracy and TPR values relative to FPR are consistently observed across various threshold settings used to define the blue regions and localized vibrational modes (see Supplementary Materials). These results reinforce our conclusion that the local structures associated with reduced thermal conductivity spatially correspond to the distribution of the low-frequency localized modes, thereby providing a clear physical interpretation of the regression results obtained from the PH-based descriptors.

**4 Conclusion**

In this study, we demonstrated that persistent homology provides a powerful and interpretable descriptor for predicting thermal conductivity in amorphous graphene. The PH-based features enabled highly accurate regression performance, revealing that structural motifs associated with low thermal conductivity arise from distorted polycyclic configurations, whereas more regular hexagonal arrangements enhance heat transport. Our inverse analysis further showed that these topological features correspond closely to low-frequency localized vibrational modes, indicating that persistent homology can capture hierarchical structural information that is typically inaccessible to conventional ring-statistics-based methods.

These findings highlight persistent homology as a promising framework for uncovering physically meaningful structure–property relationships in disordered materials. The approach opens pathways for applying topological descriptors to broader classes of amorphous and nanostructured systems, and for integration with nonlinear machine-learning models to further enhance predictive capability.

We also acknowledge several limitations. The structures analyzed here were generated through classical molecular dynamics simulations and may be influenced by the choice of interatomic potentials, simulation protocols, and finite-size effects. Experimental amorphous graphene may exhibit a broader range of structural motifs and thermal conductivities. Future work should therefore employ more realistic structure-generation methods—such as those based on machine-learned potentials—and validate the analysis through comparison with experimental data.

**Acknowledgements**

This study was supported by JST PRESTO Grant Number JPMJPR2198, JST FOREST Grant Numbers JPMJFR236Q and JPMJFR222G, MEXT KAKENHI 21H01816, 23H04470, and a grant from the Inamori Foundation. The calculations were performed using a computer facility at the Research Center for Computational Science (Okazaki, Japan, 25-IMS-C126).

Tables

Table 1 Number of hexagon and triangle structures assigned to the pairs in the red and blue regions

|  | 10K/ps | | 1000K/ps | |
| --- | --- | --- | --- | --- |
|  | hexagon | triangle | hexagon | triangle |
| Red region | 365 | 500 | 27 | 136 |
| Blue region | 65 | 413 | 130 | 329 |

Table 2 Summary of the classification metrics evaluating the spatial correspondence between the localized vibrational modes and the blue regions identified by the PH-based inverse analysis.

| Cooling rate | Accuracy | TPR | FPR |
| --- | --- | --- | --- |
| **10 K/ps** | 0.8654 | 0.7522 | 0.0938 |
| **1000 K/ps** | 0.8600 | 0.7792 | 0.1116 |

Figures and captions

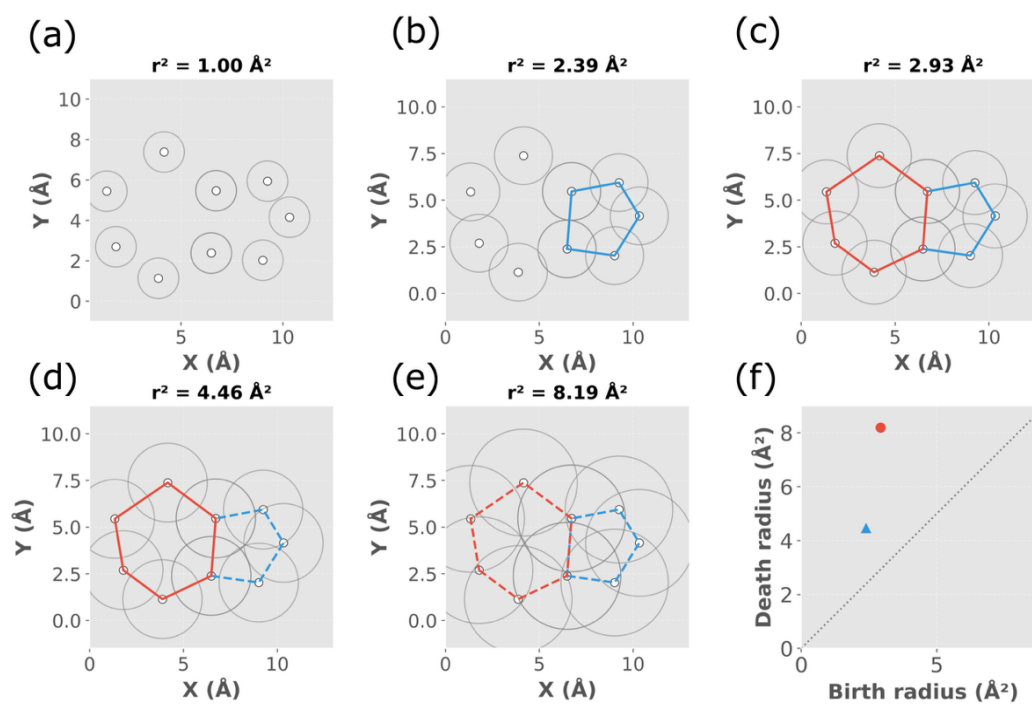

**Figure 1. Schematic illustration of persistent homology analysis based on atomic coordinates.** Virtual spheres are placed on each atom and their radii $r$ are gradually increased, during which the spheres begin to touch, connect, and eventually form and fill rings **(a–e)**. For each ring, the radius at which it appears is defined as the birth, and the radius at which it is filled is defined as the death. Plotting these pairs of radii as points yields a two-dimensional persistence diagram **(f)**. The birth and death values correspond to characteristic length scales of a structure, enabling a quantitative characterization of structural disorder and order. The squared values of birth and death radii are used to plot the persistence diagram following the usual convention.

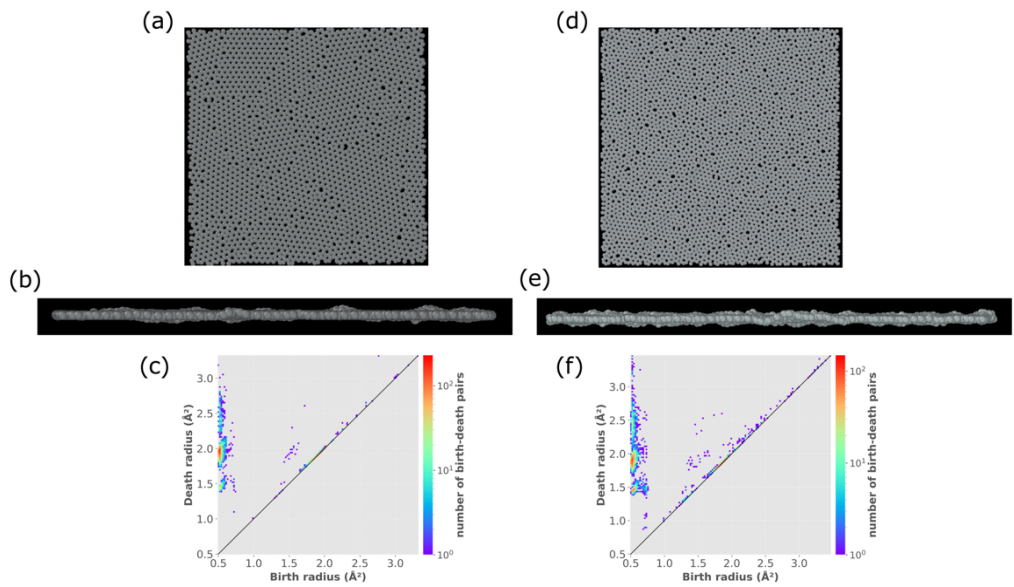

**Figure 2. Atomic structures and persistence diagrams of amorphous graphene generated at different cooling rates.**

**a, d)** Top views of amorphous graphene structures obtained at cooling rates of 10 K/ps and 1,000 K/ps, respectively. **b, e)** Corresponding side views showing the atomic thickness and surface corrugation. **c, f)** Persistence diagrams computed from the respective atomic configurations. Each point represents a birth–death pair, and the color scale indicates its multiplicity.

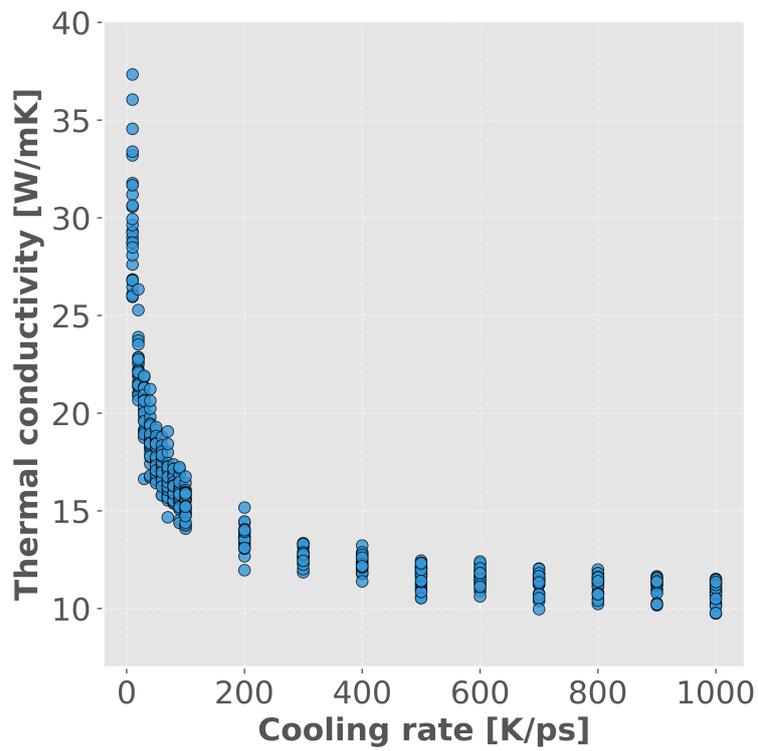

**Figure 3. Cooling-rate dependence of the thermal conductivity of amorphous graphene.** Thermal conductivity of amorphous graphene as a function of cooling rate varied from 10 to 1,000 K/ps. As the cooling rate increases, the thermal conductivity decreases sharply and then converges to an approximately constant value above about 700 K/ps.

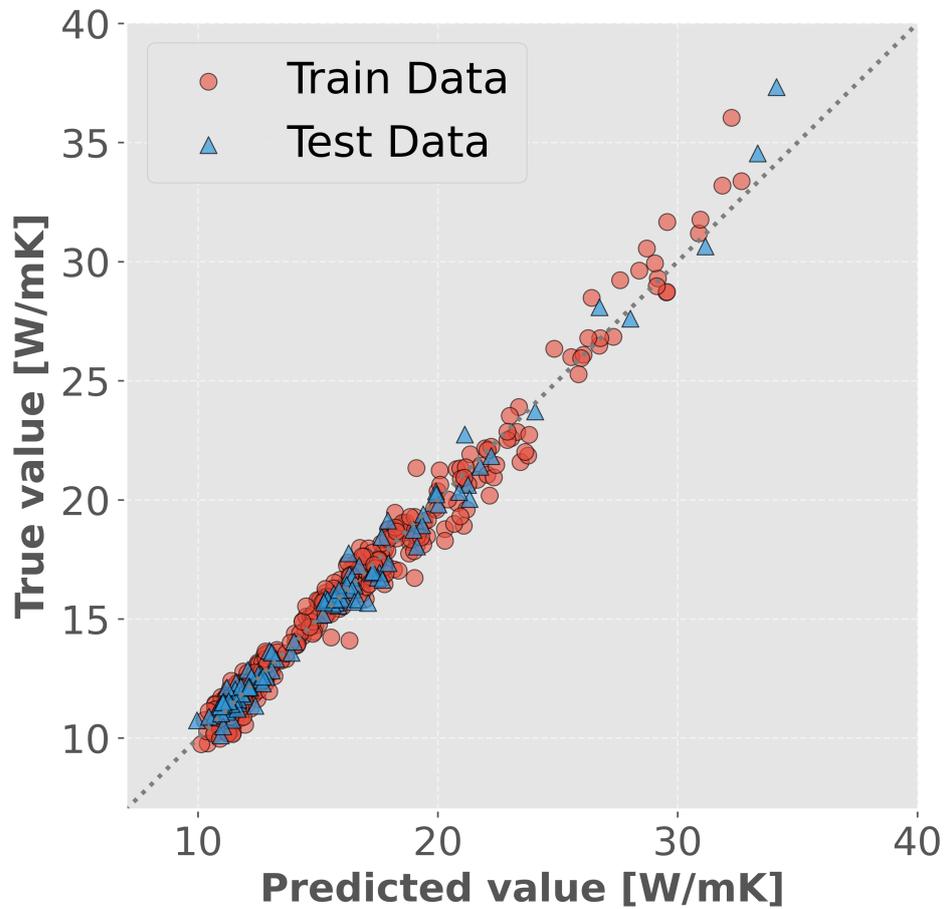

**Figure 4. Prediction accuracy of thermal conductivity using ridge regression with persistence image descriptor.**

Results of ridge regression modeling of the relationship between the descriptor obtained from persistence images and the thermal conductivity. A total of 407 data points were used for training and 102 for testing. The horizontal and vertical axes represent the predicted and true values, respectively. The coefficient of determination is $R^2 = 0.983$, and the root mean square error (RMSE) is 0.642 W m$^{-1}$ K$^{-1}$. The data points for training and test sets are shown as red circles and blue triangles, respectively.

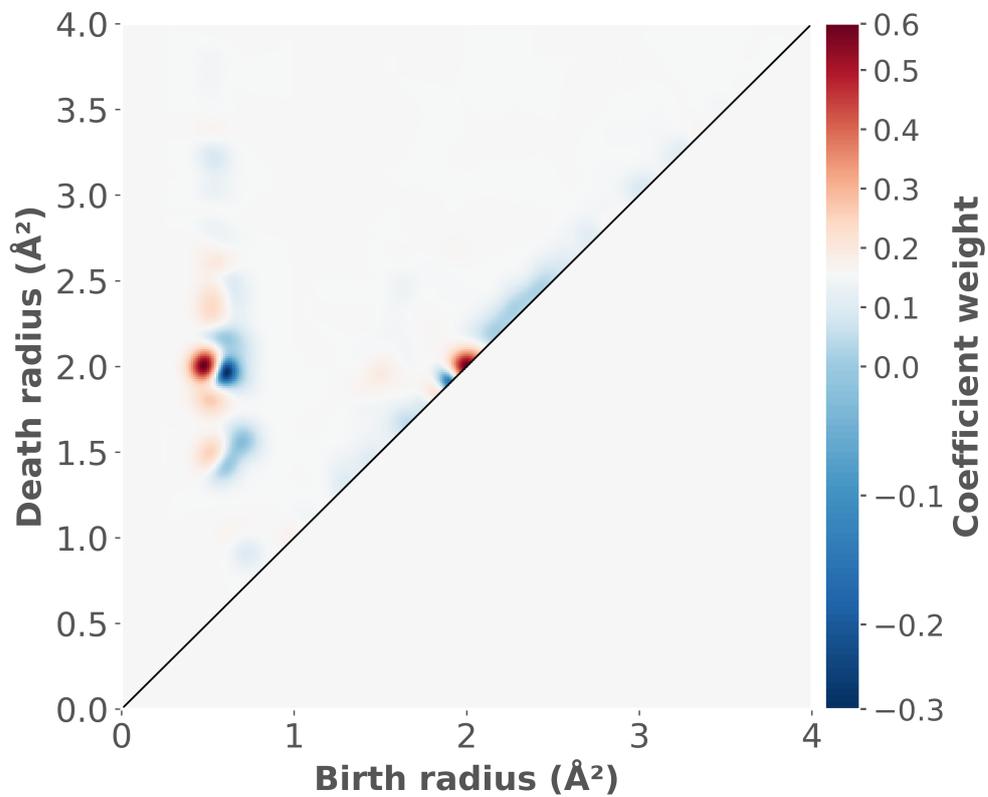

**Figure 5. Visualization of thermal conductivity contributions based on the coefficient distribution of the ridge regression model.**

The contribution of each birth–death pair to the thermal conductivity is visualized using the coefficients of the trained ridge regression model on the persistence image grid. Red indicates a positive contribution to thermal conductivity, while blue indicates a negative contribution.

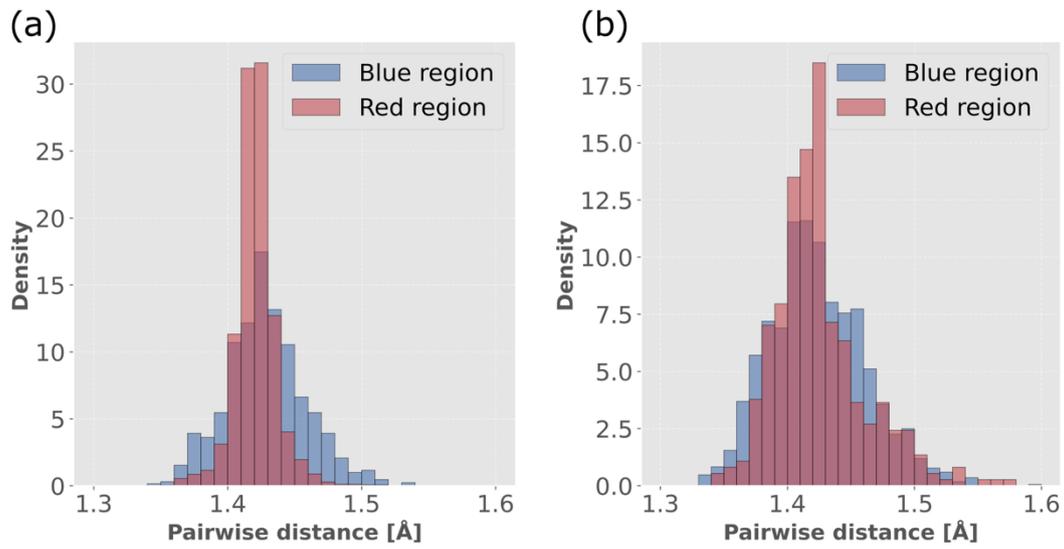

**Figure 6. Bond-length distributions of local structures contributing to thermal conductivity.**
(a) and (b) show the bond-length distributions of polygons corresponding to birth–death pairs in the PDs obtained from a-Gr structures generated at the cooling rates of 10 K/ps and 1000 K/ps, respectively. Red color corresponds to pairs in regions that contribute to increasing thermal conductivity, while blue color corresponds to those that contribute to decreasing it.

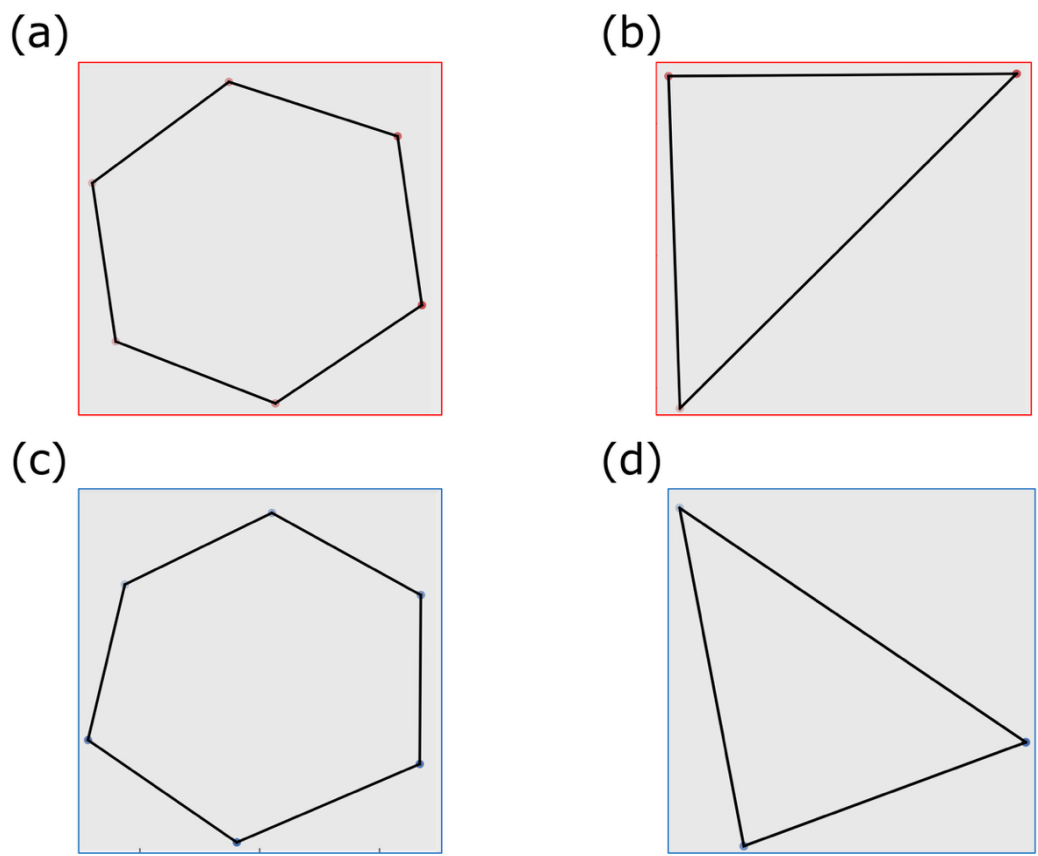

**Figure 7. Representative local structures associated with thermal conductivity.**
(a, b) Local motifs corresponding to birth–death pairs in the red region of the persistence diagram, which are correlated with higher thermal conductivity. (c, d) Local motifs corresponding to birth–death pairs in the blue region, which are correlated with lower thermal conductivity.

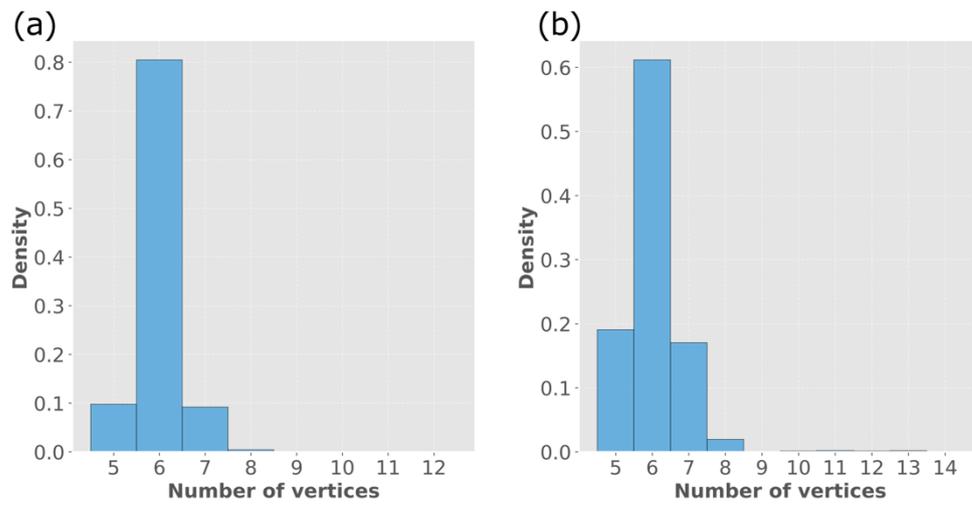

**Figure 8. Results of ring statistics analysis.**

(a) and (b) are the results of ring statistics analysis of amorphous graphene structures generated at cooling rates of 10 K/ps and 1,000 K/ps, respectively.

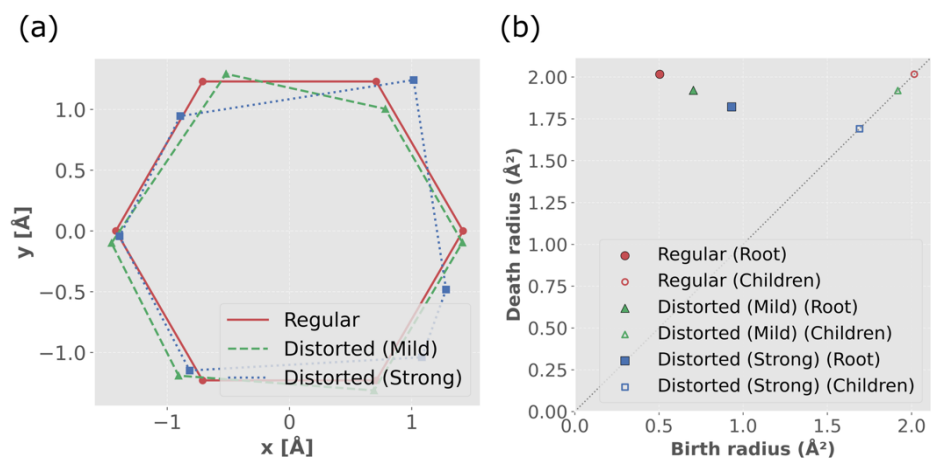

**Figure 9. Schematic illustration showing the correlation between structural disorder and the distribution of birth–death pairs in the PD.**
(a) A regular hexagonal structure with an edge length of 1.42 Å (red) and distorted hexagonal structures (green and blue) with the same perimeter but uneven edge lengths. The blue hexagon contains larger disorder compared with the green one. The procedure used to introduce disorder into each hexagon is summarized in the Supplementary Materials. (b) The corresponding PDs obtained from each structure. Circular markers represent birth–death pairs of the rings without parents (roots), while triangular markers correspond to those of subordinate rings (children).

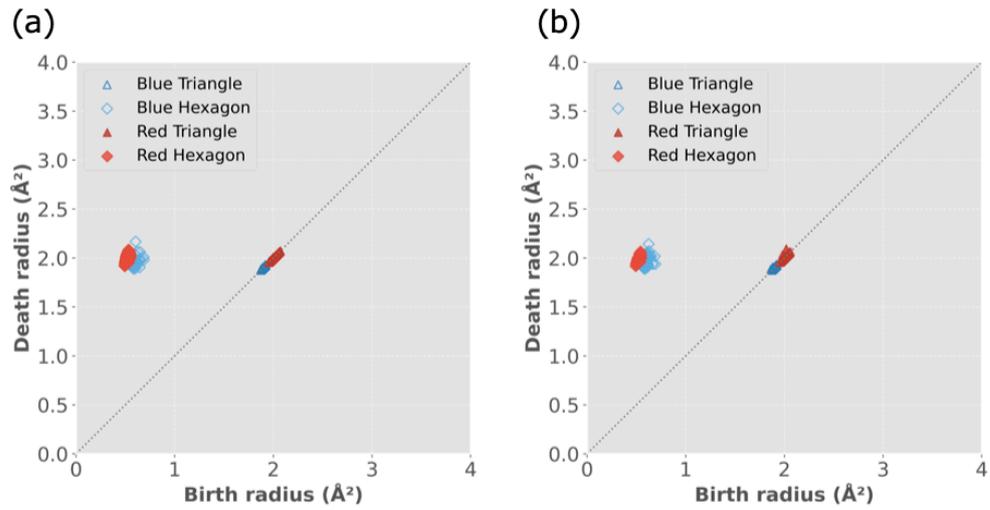

**Figure 10. Distributions of birth and death radii corresponding to hexagonal and triangular structures.**

(a) and (b) show the corresponding distributions of birth and death radii for hexagonal and triangular structures in the blue and red regions in the sample generated at the cooling rate of 10 K/ps and 1,000 K/ps, respectively.

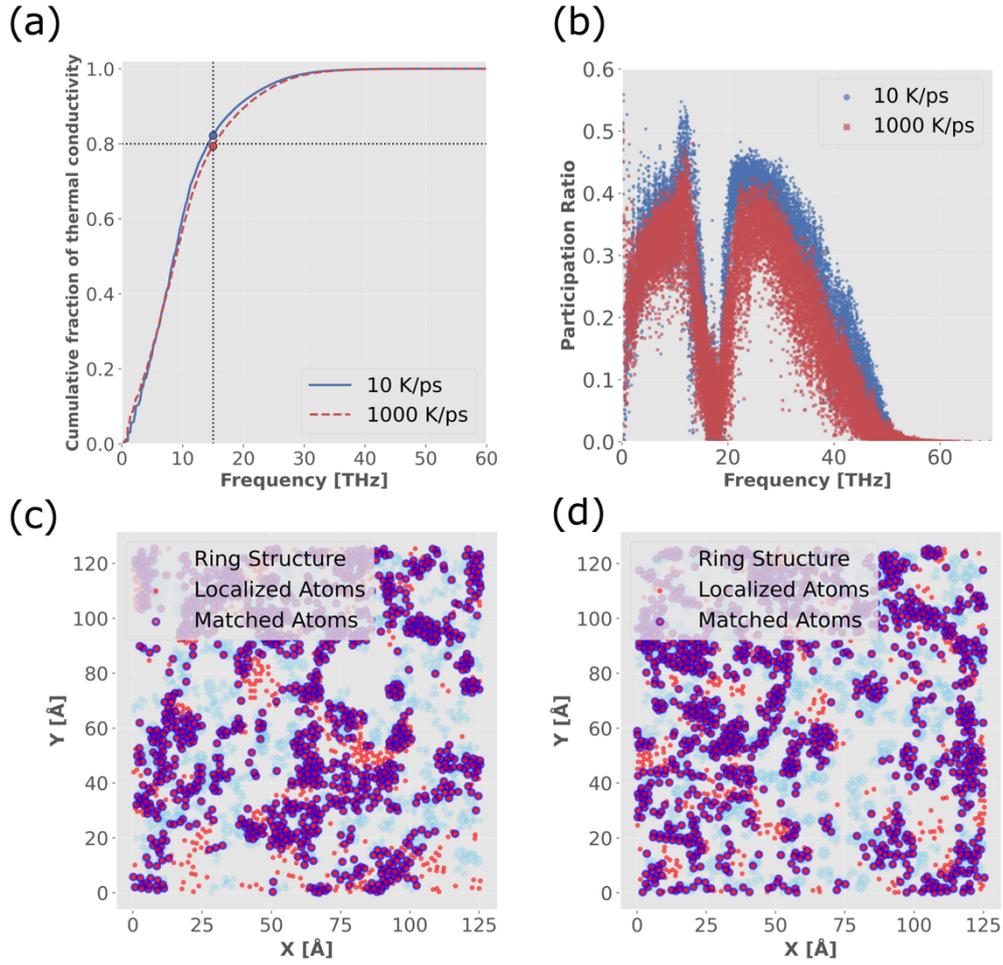

**Figure 11. Correlation between low-frequency localized vibrations and local structures contributing to reduced thermal conductivity in amorphous graphene.**

(a) Cumulative contributions to thermal conductivity as a function of frequency for cooling rates of 10 K/ps (blue) and 1000 K/ps (red). (b) Relationship between the participation ratio (PR) and vibrational frequency. (c) Spatial overlap between the low-thermal-conductivity regions (ridge regression coefficients ≤ −0.1) and low-frequency localized vibrational modes (PR ≤ 0.1) for the structure generated at a cooling rate of 10 K/ps. Atoms in the low-thermal-conductivity regions are shown by blue filled circle, those exhibiting high vibrational localization are shown by red dots, and atoms common to both are highlighted by dark blue circles. (d) Corresponding comparison for the structure generated at a cooling rate of 1000 K/ps.